%% file: main.tex
\documentclass[conference]{IEEEtran}

\usepackage{multirow}
\usepackage{booktabs, tabularx, tabularray}
\usepackage{amsmath,amssymb,amsfonts}
\usepackage{rotating}
\usepackage{adjustbox}
\usepackage{graphicx}
\usepackage{lscape}
\usepackage{algorithm}
\usepackage{algpseudocode}
\usepackage{caption, subcaption, array, cite}
\UseTblrLibrary{booktabs}
\usepackage[table]{xcolor}
\usepackage{colortbl}
\usepackage{wrapfig}
\usepackage{siunitx, threeparttable}
\usepackage{makecell}
\begin{document}

\bstctlcite{IEEEexample:BSTcontrol}
\title{VitaLLM: A Versatile and Tiny Accelerator for Mixed-Precision LLM Inference on Edge Devices}

\author{\IEEEauthorblockN{Zi-Wei Lin, and Tian-Sheuan Chang}
\IEEEauthorblockA{\textit{Institute of Electronics, National Yang Ming Chiao Tung University}
Hsinchu, Taiwan \\
}
}
\maketitle

\input{abs/abstract_en}
\input{chapters/1-Introduction}
\input{chapters/2-Compute-Core-Microarchitectures}
\input{chapters/3-System-Integration-and-Scheduling}
\input{chapters/4-Experimental-Results}
\input{chapters/5-Conclusion}
\bibliographystyle{IEEEtran}
\bibliography{bib/ieeeBSTcontrol,bib/thesis}
\end{document}

%% file: abs/abstract_en.tex
\begin{abstract}
We present \emph{VitaLLM}, a mixed-precision accelerator that enables ternary-weight large language models to run efficiently on edge devices. The design combines two compute cores—a multiplier-free \textbf{TINT} core for ternary--INT projections and a \textbf{BoothFlex} core that reuses a radix-4 Booth datapath for both INT8$\times$INT8 attention and ternary--INT—sustaining utilization without duplicating arrays. A predictive sparse attention mechanism employs a leading-one (LO) surrogate with a comparison-free top-$K$ selector to prune key/value (KV) fetches by roughly $1{-}K/M$ for $M$ cached tokens, confining exact attention to $K$ candidates. System-level integration uses head-level pipelining and an absmax-based quantization barrier to standardize cross-core interfaces and overlap nonlinear reductions with linear tiles. A 16\,nm silicon prototype at 1\,GHz/0.8\,V achieves 72.46 tokens/s in decode and 0.88 s prefill (64 tokens) within 0.214 mm² and 120 KB on-chip memory, while reducing KV traffic and improving utilization in ablations. These results demonstrate practical BitNet b1.58 (3B) inference on edge-class platforms and provide a compact blueprint for future mixed-precision LLM accelerators.
\end{abstract}

%% file: chapters/1-Introduction.tex
\section{Introduction}

Large language models (LLMs) have revolutionized natural language understanding, but their growing scale stresses computation, memory, and energy—especially on edge platforms that require always-on, low-latency inference under tight power and capacity budgets. Meeting these constraints demands model compression and hardware co-design, particularly below 8-bit precision. 

\begin{figure}[htbp!]
    \centering
    \vspace{-2mm}
    \includegraphics[width=0.75\linewidth]{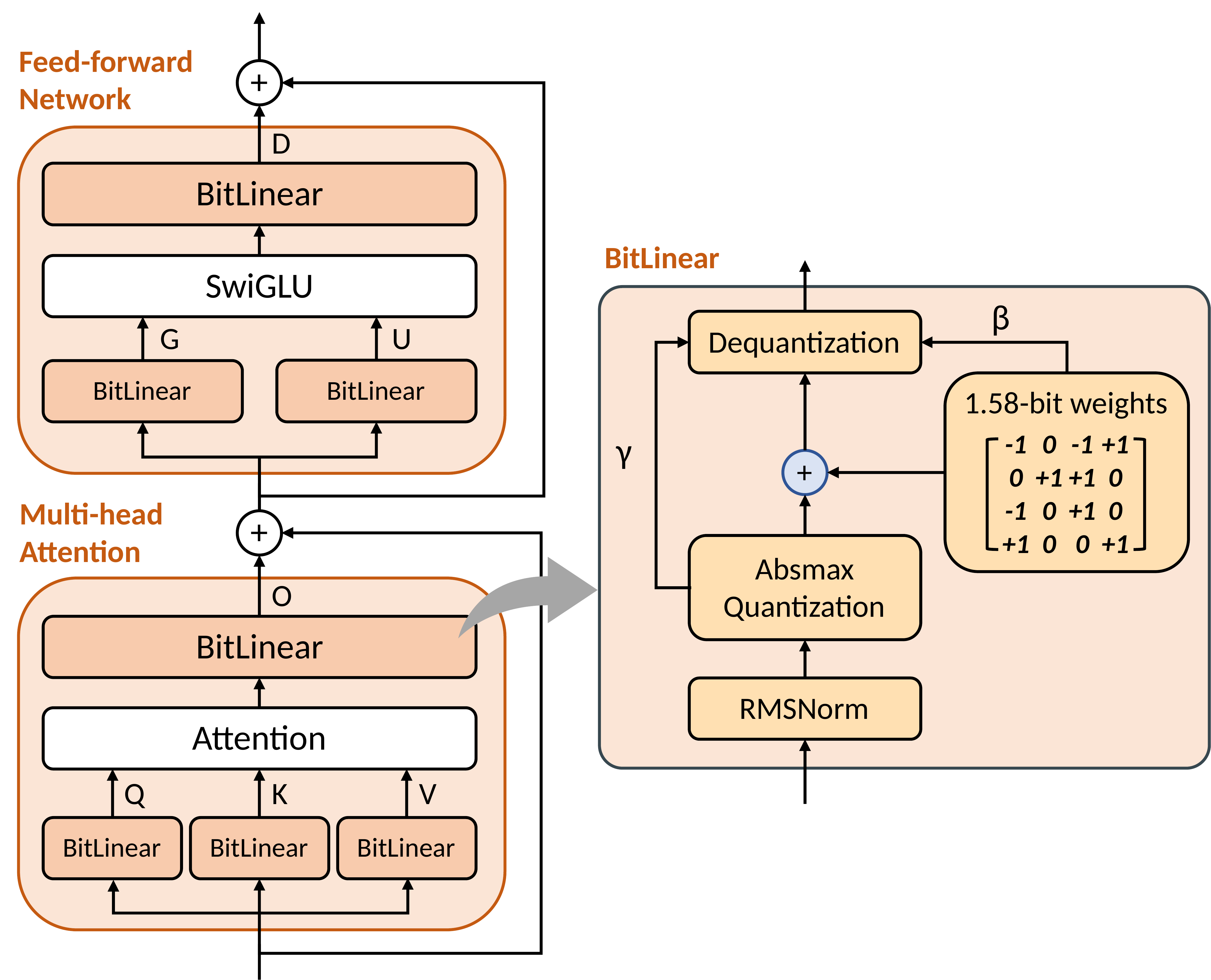}
    \caption{BitNet b1.58 architecture with BitLinear replacing all weight projection layers.}
    \label{fig:bitnet_diagram}
    \vspace{-2mm}
\end{figure}

BitNet b1.58~\cite{bitnet158,bitnet2b4t} advances this goal by using ternary weights $\{-1,0,+1\}$ with 8-bit activations. Trained with quantization-aware methods, BitNet b1.58 achieves competitive accuracy from the 3B-parameter scale while reducing memory (up to $7\times$ reduction) and computation. In BitNet, BitLinear layers replace conventional linear projections in both attention and feedforward modules (see Fig.~\ref{fig:bitnet_diagram}).

Specialized hardware is required to realize these algorithmic gains in practice. Prior ternary LLM accelerators include Slim-Llama (ASIC)~\cite{slimllama}, TerEffic (FPGA)~\cite{tereffic} and TeLLMe (edge FPGA)~\cite{tellme}. Nonetheless, three challenges remain for edge deployment: (i) poor array utilization between ternary and INT8 attention, (ii) high KV-cache bandwidth from random access, and (iii) complex mixed-precision/nonlinearity interfaces.

We present \emph{VitaLLM}, an edge-oriented mixed-precision accelerator that addresses these challenges with three contributions: 
\begin{itemize}
  \item \textbf{Dual-core compute.} A multiplier-free \textbf{TINT} core for all ternary--INT projections and a \textbf{BoothFlex} core that executes both INT8$\times$INT8 (attention) and ternary--INT on a shared radix-4 Booth datapath, improving utilization without duplicating arrays.
  \item \textbf{Predictive sparse attention.} A leading-one (LO) surrogate with a comparison-free top-$K$ selector that reduces KV fetches by roughly $1-\!K/M$ (for $M$ cached tokens) and confines exact attention to the $K$ candidates—no retraining required.
  \item \textbf{System integration.} Head-level pipelining that consumes Q/K/V immediately per head, plus an absmax-based quantization barrier that decouples nonlinear reductions from scaling to provide a uniform \texttt{(integer vector, single scale)} interface across cores.
\end{itemize}

The remainder of this paper is organized as follows: Section~\ref{sec:compute-core} describes the TINT and BoothFlex microarchitectures. Section~\ref{sec:system} presents system integration. Section~\ref{sec:impl_eval} reports implementation and measurements. Section~\ref{sec:conclusion} concludes.

%% file: chapters/2-Compute-Core-Microarchitectures.tex
\section{Compute Core Microarchitectures}
\label{sec:compute-core}

We employ two multiplier-light compute arrays that share on-chip buffers and a common output-stationary (OS) dataflow. The \textbf{TINT} core handles all ternary--INT projections, while \textbf{BoothFlex} reuses a radix-4 Booth pipeline to execute both INT8$\times$INT8 attention and ternary--INT without duplicating arrays. This co-design sustains hardware utilization while containing area and memory traffic.

\subsection{TINT Core}
\label{ssec:tint}

\begin{figure}[htbp!]
  \centering
  \includegraphics[width=0.75\linewidth]{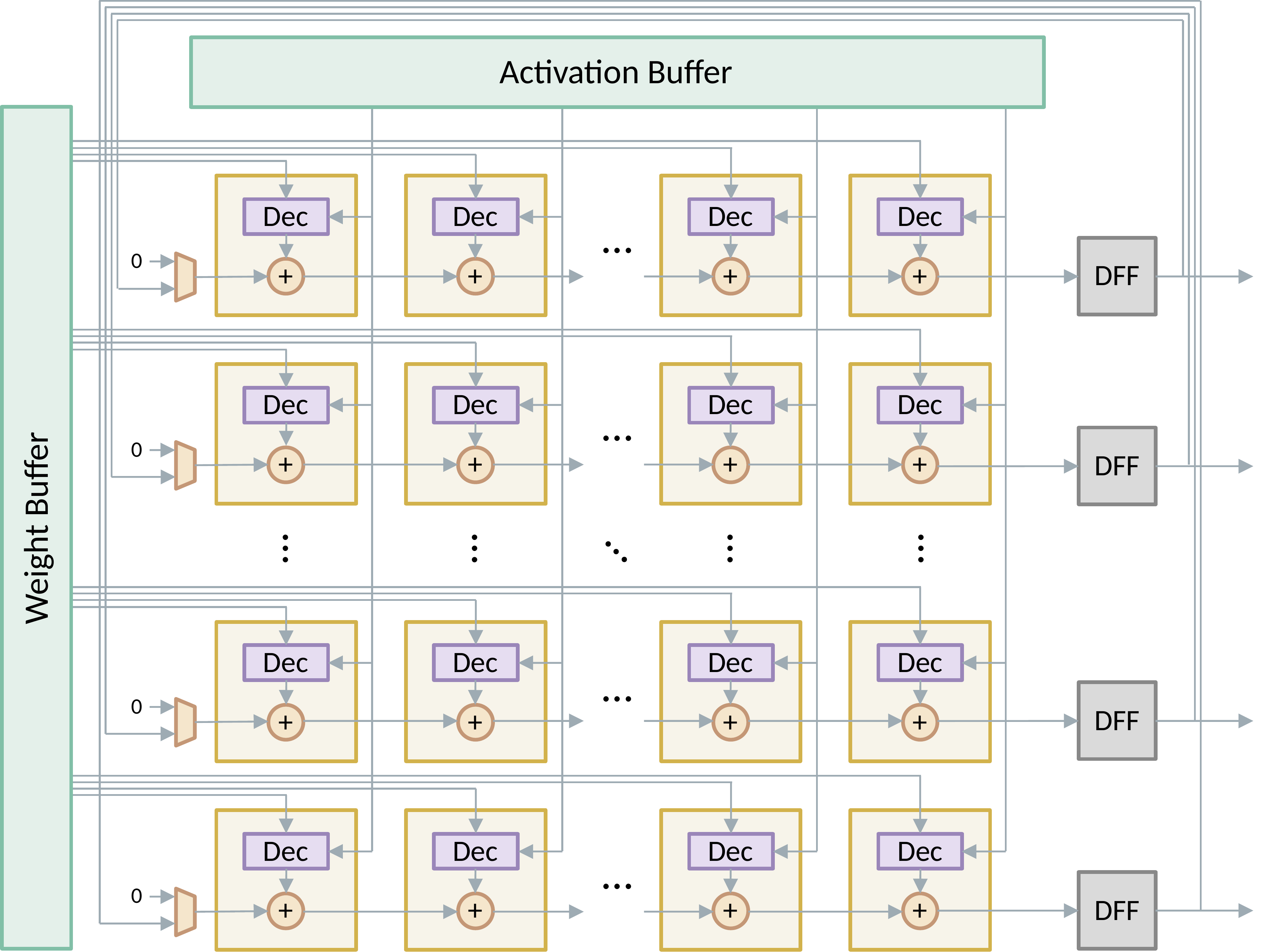}
  \vspace{-2mm}
  \captionsetup{skip=8pt}
  \caption{TINT core: $8{\times}8$ multiplier-free PE array for ternary–INT with output-stationary mapping and tiled outputs.}
  \label{fig:tint_core}
  \vspace{-2mm}
\end{figure}

TINT (Fig.~\ref{fig:tint_core}) accelerates ternary--INT dot products with a multiplier-free $8{\times}8$ PE array and OS mapping. Each PE receives an activation $a$ and a ternary weight $w\!\in\!\{-1,0,+1\}$, and updates a local accumulator using a lightweight selector:
\[
y \leftarrow y + \mathrm{sel}(w,a),\quad \mathrm{sel}(w,a)\in\{0,\,+a,\,-a\}.
\]
The selector is implemented as a small mux with sign inversion driven by a ternary decoder, avoiding general-purpose multipliers. Per cycle, the array broadcasts eight INT activations across columns while all $8{\times}8$ PEs consume 64 ternary weights, sustaining 64 ternary-INT select-accumulates per cycle. Outputs are emitted in fixed-size tiles while partial sums remain on-chip, minimizing buffer traffic and toggling. A compact control path streams packed ternary codes and INT activations, enabling high throughput for Q/K/V, $W_O$, and FFN projections on a ternary-arithmetic datapath.

\subsection{BoothFlex Core}
\label{ssec:boothflex}

\begin{figure}[htbp!]
  \centering
  \includegraphics[width=0.86\linewidth]{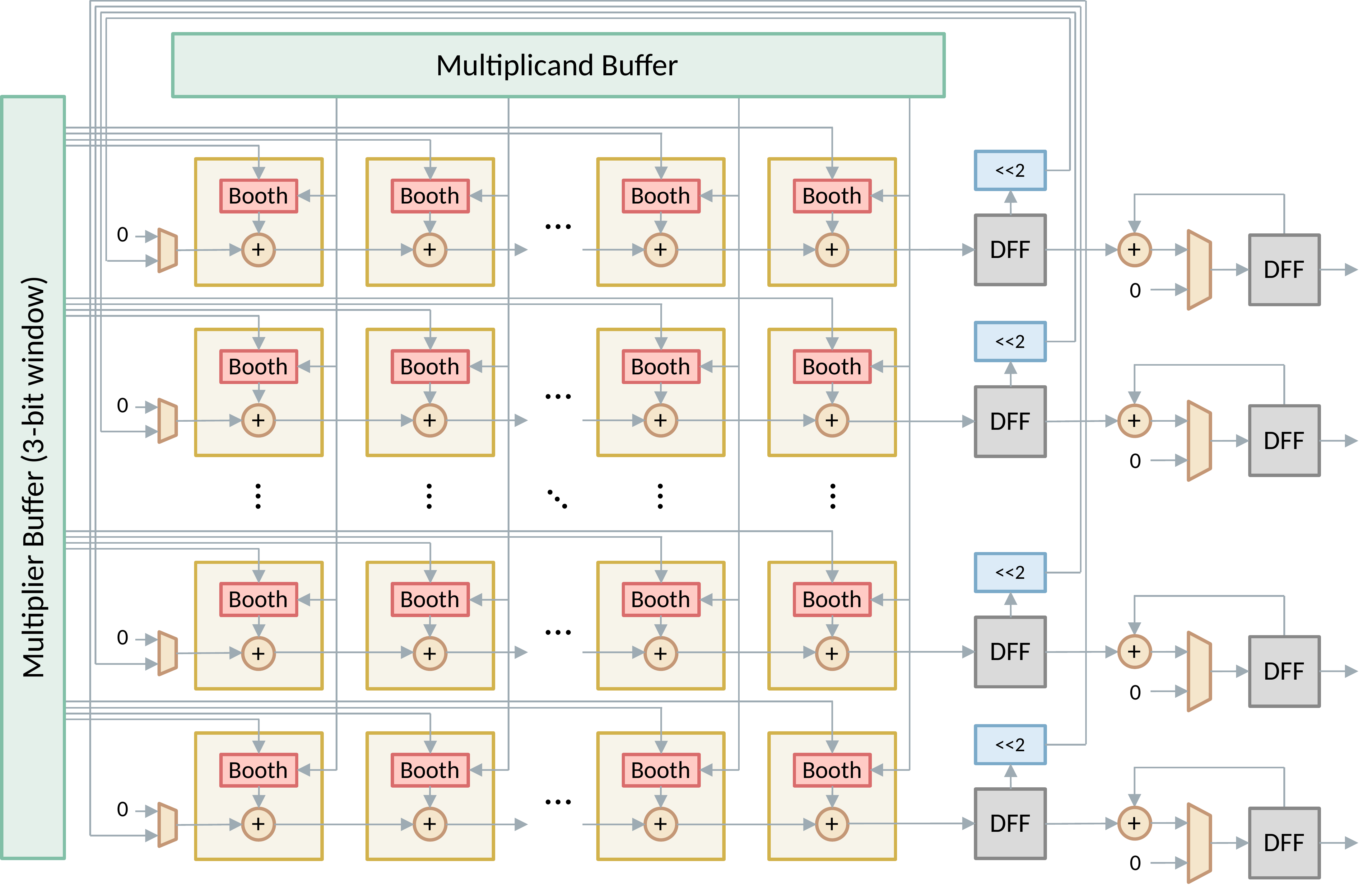}
  \vspace{-2mm}
  \captionsetup{skip=6pt}
  \caption{BoothFlex core: shared radix-4 Booth array supporting ternary–INT (1 iteration via a single 3-bit window) and INT8$\times$INT8 (5 iterations with row-wise ``$\ll2$'' shifts).}
  \label{fig:boothflex_arch}
  \vspace{-2mm}
\end{figure}

Transformer attention requires INT8$\times$INT8 products in addition to ternary–INT. BoothFlex reuses one radix-4 Booth datapath for both modes (Fig.~\ref{fig:boothflex_arch}). The recoder scans overlapping 3-bit windows $\{y_{2i+1},y_{2i},y_{2i-1}\}$ (with $y_{-1}{=}0$) to produce $r_i\!\in\!\{-2,-1,0,+1,+2\}$. For a $b$-bit multiplier, the iteration count is \(N=\left\lceil \frac{b+1}{2}\right\rceil\). Thus, INT8$\times$INT8 requires $b{=}8\Rightarrow N{=}5$ cycles. For ternary–INT, we zero-pad the 2-bit ternary code to one 3-bit window ($-1{:}11\!\Rightarrow\!110,\;0{:}00\!\Rightarrow\!000,\;+1{:}01\!\Rightarrow\!010$), yielding $N{=}1$. Thus ternary and INT8 operands share the same Booth pipeline without extra arithmetic units.

At the PE level, computation is bit-serial: one Booth iteration per cycle with row accumulators updated as
\[
\mathrm{PS}_i \leftarrow (\mathrm{PS}_{i-1}\ll 2) + \mathrm{PP}_i ,
\]
where $\mathrm{PP}_i$ sums partial products selected by $r_i$. INT8$\times$INT8 spans five cycles; ternary–INT collapses to a single cycle. As with TINT, OS mapping keeps partial sums local and outputs are emitted in 8-element chunks. This schedule trades small latency for sizable area/power savings, while the head-level pipeline (Sec.~\ref{subsec:head_pipeline}) later maintains high utilization.

%% file: chapters/3-System-Integration-and-Scheduling.tex
\section{System Integration and Scheduling}
\label{sec:system}

VitaLLM integrates the TINT and BoothFlex cores under a shared output-stationary (OS) schedule and lightweight control network. The schedule keeps arrays utilized by (i) pruning attention candidates before costly KV fetch, (ii) pipelining work at the \emph{head} granularity, and (iii) standardizing mixed-precision interfaces via an absmax-based quantization barrier.

\subsection{Predictive Sparse Attention}
\label{subsec:pred-sparse}

We reduce key/value (KV) traffic and attention MACs by screening candidate keys using a multiplier-free \textbf{Leading-One Prediction (LOP)} surrogate that correlates with the true dot product.

\begin{figure}[t!]
  \centering
  \includegraphics[width=0.88\linewidth]{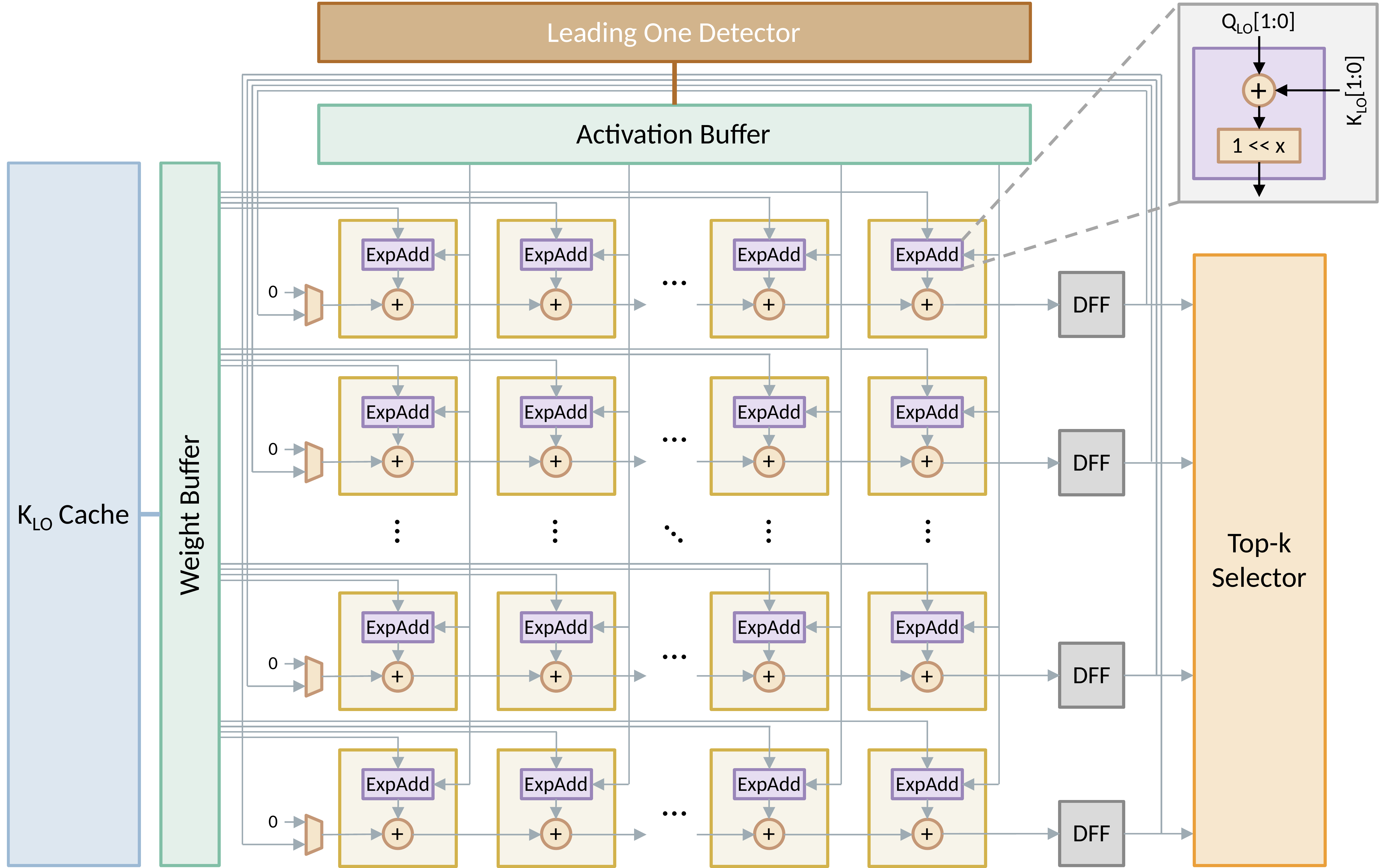}
  \vspace{-2mm}
  \captionsetup{skip=8pt}
  \caption{LOP array with a comparison-free top-$K$ selector: $(\mathrm{LO},\mathrm{sgn})$ features feed \emph{ExpAdd} PEs; bucketized scores stream $K$ indices without pairwise compares.}
  \label{fig:lop_arch}
  \vspace{-2mm}
\end{figure}

Let $q,k\!\in\!\mathbb{Z}^d$ be INT8 query and key vectors. Define
\[
\mathrm{LO}(x)=\Big\lfloor \log_2 |x| \Big\rfloor,\qquad 
\operatorname{sgn}(x)\in\{-1,+1\}.
\]
The surrogate score is
\begin{equation}
\hat{s}(q,k)=\sum_{i=1}^{d}\operatorname{sgn}(q_i)\operatorname{sgn}(k_i)\;2^{\,\mathrm{LO}(q_i)+\mathrm{LO}(k_i)}.
\label{eq:lop}
\end{equation}
Each term reduces to a barrel-shifted 1, enabling a compact \emph{ExpAdd} PE that streams $(q,k)$ once to accumulate~\eqref{eq:lop}. A streaming top-$K$ (or threshold) selector emits only the highest-scoring indices for exact attention, while all others are skipped (Fig.~\ref{fig:lop_arch}). This maintains attention quality and converts random KV probes into short, contiguous reads that the memory system can serve efficiently.

To achieve fixed-$K$ selection with predictable throughput and small area, we adopt a \emph{comparison-free}, $k$-degree parallel design inspired by~\cite{ray2023kdegree}. Scores are bucketized into discrete ranges; a high-to-low prefix scan locates the cut bin where the cumulative count first reaches $K$; then $k$-wide priority encoders emit up to $k$ indices per cycle, avoiding wide comparator trees and matching the streaming access pattern to cached $K$.

At the system level: (i) \textit{KV traffic.} Only keys in the kept set $\mathcal{C}$ are fetched for exact $qK_{\mathcal{C}}^{\top}$ and the subsequent $\mathbf{SV}$, so average KV fetches scale with $K$ rather than the cache length $M$. (ii) \textit{Arithmetic.} Computing~\eqref{eq:lop} replaces dot-product multiplications with shifts and additions, and exact INT8$\times$INT8 attention is performed only for $\mathcal{C}$, reducing compute roughly with $K/M$ and simplifying the head-interleaved pipeline in Sec.~\ref{subsec:head_pipeline}, all without retraining.

\subsection{Head-Level Pipelining}
\label{subsec:head_pipeline}

\begin{figure}[t!]
  \centering
  \includegraphics[width=0.98\linewidth]{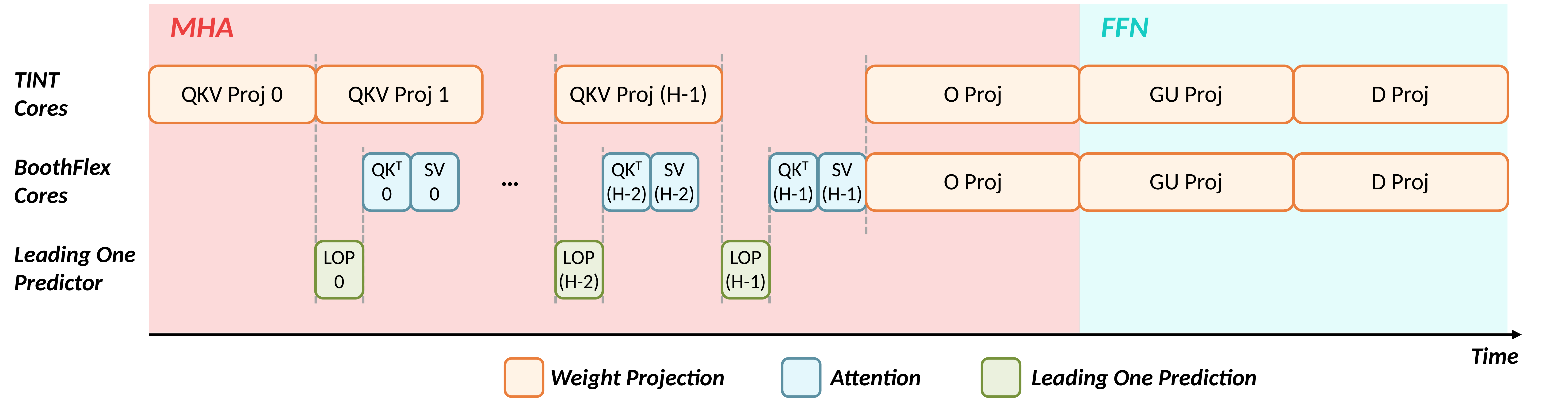}
  \vspace{-2mm}
  \captionsetup{skip=8pt}
  \caption{Head-level pipelining: TINT finishes Q/K/V for head $h$ and BoothFlex runs its attention; TINT proceeds to $h{+}1$. After MHA, both cores perform ternary–INT for $W_O$ and FFN.}
  \label{fig:head_pipeline}
  \vspace{-2mm}
\end{figure}

To balance compute and memory traffic across the two cores, we organize multi-head attention as a \emph{two-stage stream} at head granularity (Fig.~\ref{fig:head_pipeline}). A conventional schedule computes Q/K/V for \emph{all} heads and materializes $\{\mathbf Q_h,\mathbf K_h,\mathbf V_h\}_{h=0}^{H-1}$ before starting attention—incurring an extra round of writes and later re-reads. Instead, we use a one-head–offset pipeline: as soon as head $h$ finishes Q/K/V on \textbf{TINT}, the same head immediately enters attention on \textbf{BoothFlex} using LOP-derived top-$K$ indices (Sec.~\ref{subsec:pred-sparse}), while \textbf{TINT} advances to Q/K/V for head $h{+}1$. Q/K/V are consumed promptly, avoiding bulk intermediate storage and reducing inter-stage traffic. See Fig.~\ref{fig:head_pipeline}.

This schedule also exploits \textbf{BoothFlex} dual mode. During MHA, BoothFlex runs INT8$\times$INT8 for $\mathbf{QK}^{\top}$ and $\mathbf{SV}$. After all heads finish and outputs are concatenated, \emph{both cores} switch to ternary–INT and \emph{cooperate} on $W_O$ and FFN (up, gate, down). The result: no Q/K/V buffering, attention begins as soon as each head is ready, and utilization stays high as BoothFlex switches from INT8$\times$INT8 to ternary–INT after MHA.

\subsection{Nonlinear Decoupling and Absmax-Based Quantization Barrier}
\label{subsec:absmax_barrier}

\begin{figure}[htbp!]
  \centering
  \includegraphics[width=0.88\linewidth]{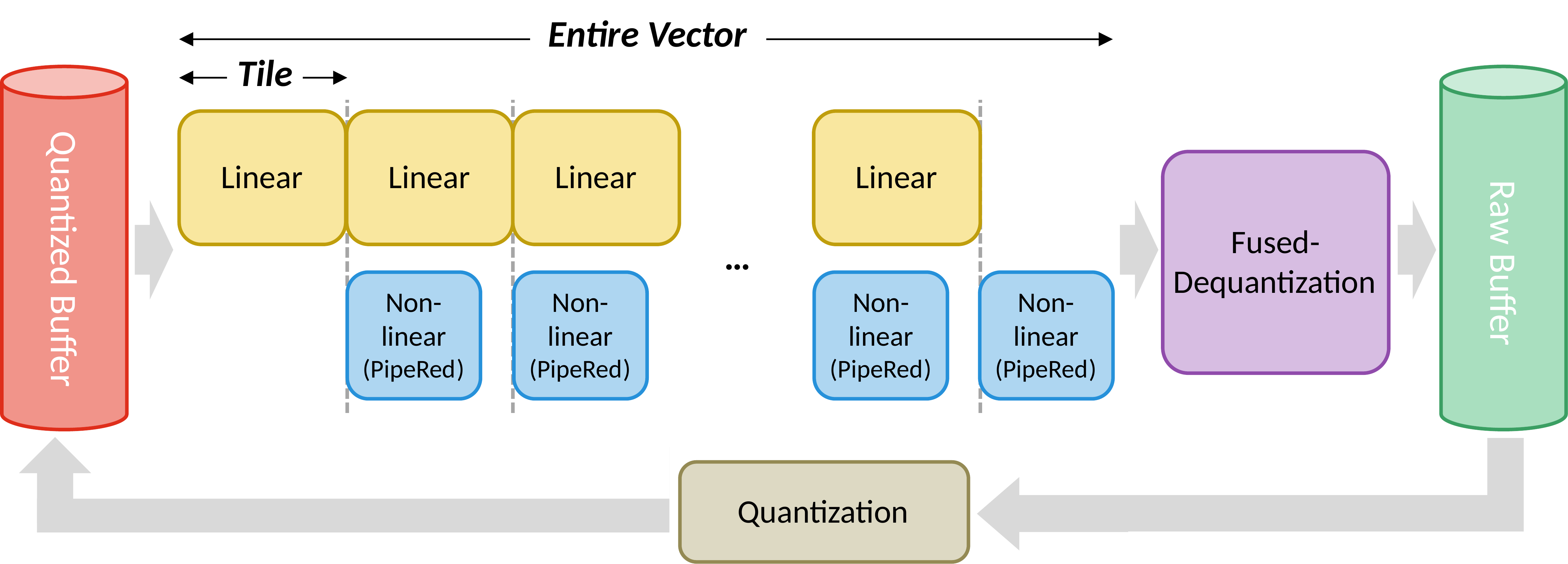}
  \vspace{-2mm}
  \captionsetup{skip=8pt}
  \caption{Absmax-based quantization barrier: reductions overlap with linear tiles; quantization is issued once per vector, with dequantization fused at the consumer}
  \label{fig:quant_barrier}
  \vspace{-2mm}
\end{figure}

Unlike fully fused flows that perform $\mathbf{QK}^{\top}\!\rightarrow\!\mathrm{softmax}\!\rightarrow\!\mathbf{S}\mathbf{V}$ inside one array~\cite{lin2025systolicattention}, VitaLLM crosses a \emph{quantized} interface between two heterogeneous cores. To keep the stream, we decouple \emph{data-dependent reductions} from \emph{scaling/normalization}. As the upstream linear kernel (e.g., QK, projections) emits tiles, a small scaling unit multiplies each tile by its activation scale and feeds a parallel reduction pipeline: \emph{running max and sum of exponentials} for softmax, or \emph{sum of squares} for RMSNorm. When the final tile of the vector arrives, these reductions are complete; we then perform the normalization over the full vector, compute a single per-vector scale, quantize once, and forward the \texttt{(integer vector, scale)} pair. The downstream linear kernel runs entirely in the integer domain and needs only one \emph{output-side} dequantization before its following nonlinearity. (See Fig.~\ref{fig:quant_barrier}.)

For our 3B BitNet b1.58 setting, inter-stage scaling uses per-vector \emph{absmax} $\alpha=\max_i |x_i|$. Because $\alpha$ itself is a vector-wide reduction, it naturally serves as the \emph{synchronization barrier}: quantization can occur only after the reduction completes. We exploit this to modularize the flow—reductions overlap with vector production, quantization fires once per vector, and the same reduction module is time-multiplexed across operators. The result is a latency-hidden, vector-granular pipeline that fits our head-level streaming schedule without feedback paths or multi-ported SRAMs (Fig.~\ref{fig:quant_barrier}).

%% file: chapters/4-Experimental-Results.tex
\section{Implementation and Experimental Evaluation}
\label{sec:impl_eval}

\subsection{Overall Architecture}
\label{subsec:overall_arch}

\begin{figure}[htbp!]
  \centering
  \includegraphics[width=0.82\linewidth]{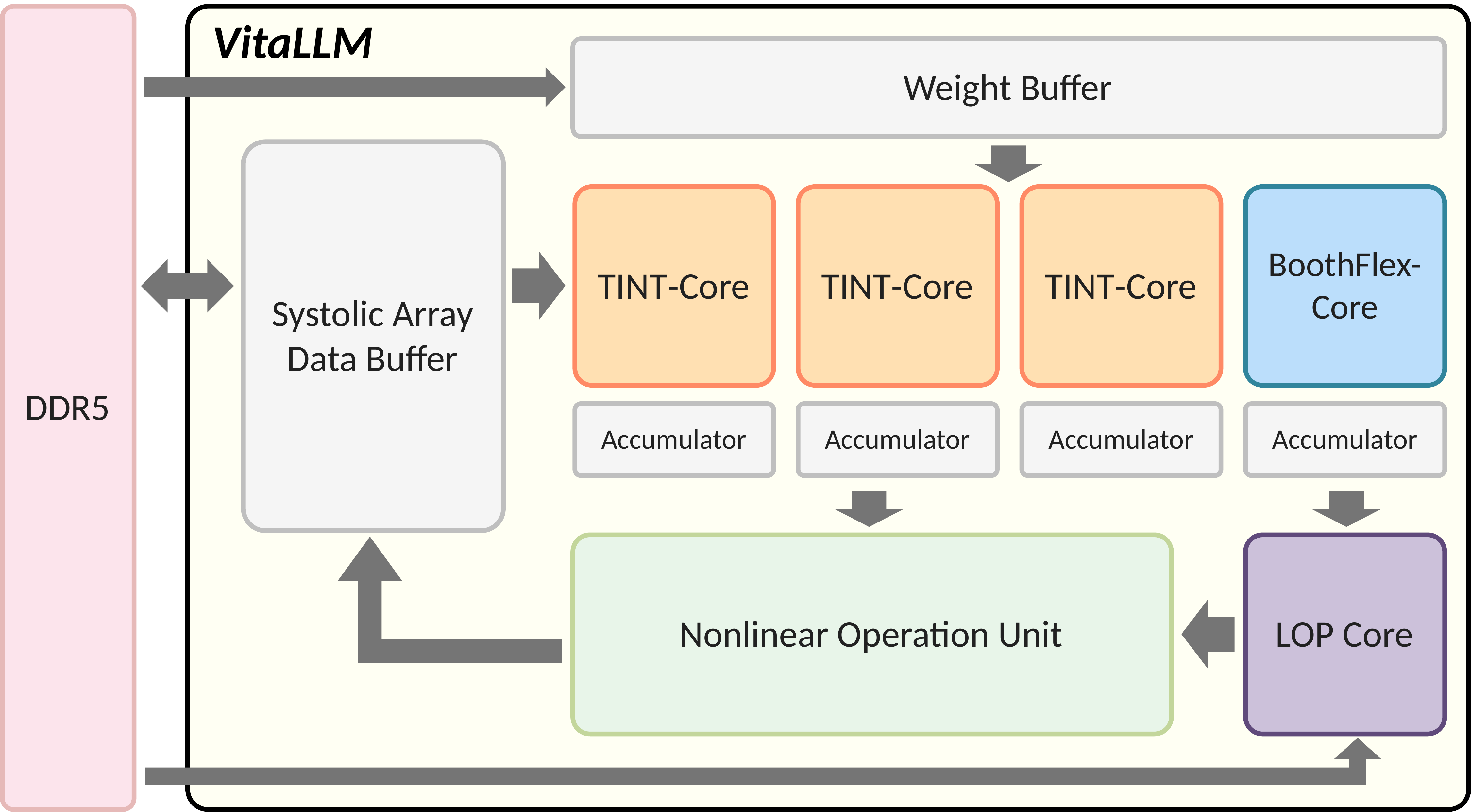}
  \vspace{-2mm}
  \captionsetup{skip=12pt}
  \caption{Overall architecture. TINT and BoothFlex arrays share on-chip buffers; LOP gates KV fetch with top-$K$ indices, while the nonlinear unit implements softmax/RMSNorm.}
  \label{fig:overall_arch}
  \vspace{-2mm}
\end{figure}

Fig.~\ref{fig:overall_arch} shows the top level. The design instantiates three \textbf{TINT} cores, one \textbf{BoothFlex} core, a lightweight \textbf{LOP} unit, a nonlinear block, and on-chip buffers. Weights and the KV cache reside in off-chip DDR; tiles are DMA mapped to on-chip buffers and executed with an output-stationary mapping so partial sums stay in local accumulators.

LOP sits on the fast path between TINT and memory. When a head’s Q/K/V tiles finish on TINT, the query features $(\mathrm{LO},\mathrm{sgn})$ are compared to cached key features to form a top-$K$ set; only those candidate blocks are requested from the KV cache for exact attention. This reduces KV traffic by roughly $1{-}K/M$ for $M$ cached tokens, and confines INT8$\times$INT8 work to the selected keys (Sec.~\ref{subsec:pred-sparse}).

Buffers are bank-shared with fixed ports (TINT$\rightarrow$A, BoothFlex$\rightarrow$B) and a small credit counter; a tile dispatcher issues ready tiles. After MHA, both cores switch to ternary–INT and cooperatively process $W_O$/FFN at the same granularity.

The RTL targets 16\,nm CMOS at 1\,GHz/0.8\,V with clock and operand gating to suppress toggling. Detailed PPA, throughput, and ablations are reported in Sec.~\ref{sec:impl_eval_results}.

\subsection{Experimental Results}
\label{sec:impl_eval_results}

We evaluate \textbf{VitaLLM} on the silicon prototype in Sec.~\ref{subsec:overall_arch} (TSMC 16\,nm, 1\,GHz, 0.8\,V).

\medskip
\noindent\textit{End-to-end comparison.}
Prior ternary accelerators make different trade-offs: ASIC \textbf{Slim-Llama} emphasizes prefill with heavy compute, which can leave the hardware underutilized in bandwidth-limited decode~\cite{slimllama}; FPGA designs (\textbf{TerEffic}, \textbf{TeLLMe}) spread resources more evenly across prefill/decode~\cite{tereffic,tellme}. \textbf{VitaLLM} brings this \emph{balanced} approach to ASIC and further sustains utilization via head-level streaming plus an LOP-driven top-$K$ gate that trims KV fetches before exact attention (Sec.~\ref{subsec:pred-sparse}). For edge settings with moderate sequence lengths, modest prefill parallelism is acceptable and keeps both cores busy.

Table~\ref{tab:sota} summarizes the results: our 16\,nm prototype achieves \textbf{72.46\,tokens/s} decode and \textbf{0.88\,s} prefill (64 tokens) within \textbf{0.214\,mm$^2$} and \textbf{120\,KB} on-chip memory.

\begin{table}[htbp!]
\centering
\footnotesize
\setlength{\tabcolsep}{3pt}
\renewcommand{\arraystretch}{1.15}
\begin{threeparttable}
\caption{Comprehensive Comparison of Ternary LLM Accelerator Designs}
\label{tab:sota}
\begin{tabular}{@{} l c c c c @{}}
\toprule
\textbf{Metric} & \textbf{VitaLLM} & \textbf{Slim-Llama~\cite{slimllama}} & \textbf{TerEffic~\cite{tereffic}} & \textbf{TeLLMe~\cite{tellme}} \\
\midrule
Platform            & ASIC 16\,nm     & ASIC 28\,nm        & FPGA U280   & FPGA KV260 \\
Freq.(MHz)     & 1000            & 25--200            & 150         & 250 \\
Voltage(V)         & 0.8             & 0.58--1.0          & --          & -- \\
On-chip Mem.        & 120\,KB         & 500\,KB            & 42\,MB      & BRAMs \\
Area(mm\textsuperscript{2}) & 0.214   & 20.25             & --          & -- \\
Power(mW)          & 59.12           & 82.07\tnote{‡}     & 31.8/46.2K  & 6.72K \\
Params              & 3B              & 3B                 & 370M--2.7B  & $\sim$0.7B \\
Prefill Time(s)    & 0.88\tnote{†}   & 0.635\tnote{*}     & --          & 0.55\tnote{†} \\
\makecell[l]{Throughput\\(tokens/s)}  & 72.46         & --                 & 727         & 9.51 \\
\bottomrule
\end{tabular}

\begin{tablenotes}[para,flushleft]
\footnotesize
\item[] * 1024 input tokens; † 64 input tokens; ‡ Measured at 200\,MHz, 1.0\,V.
\end{tablenotes}
\end{threeparttable}
\end{table}

\medskip
\noindent\textit{Ablation studies.}
We isolate the contribution of each system component using system evaluation and the figures below.

\begin{figure}[htbp!]
  \centering
  \includegraphics[width=0.46\linewidth]{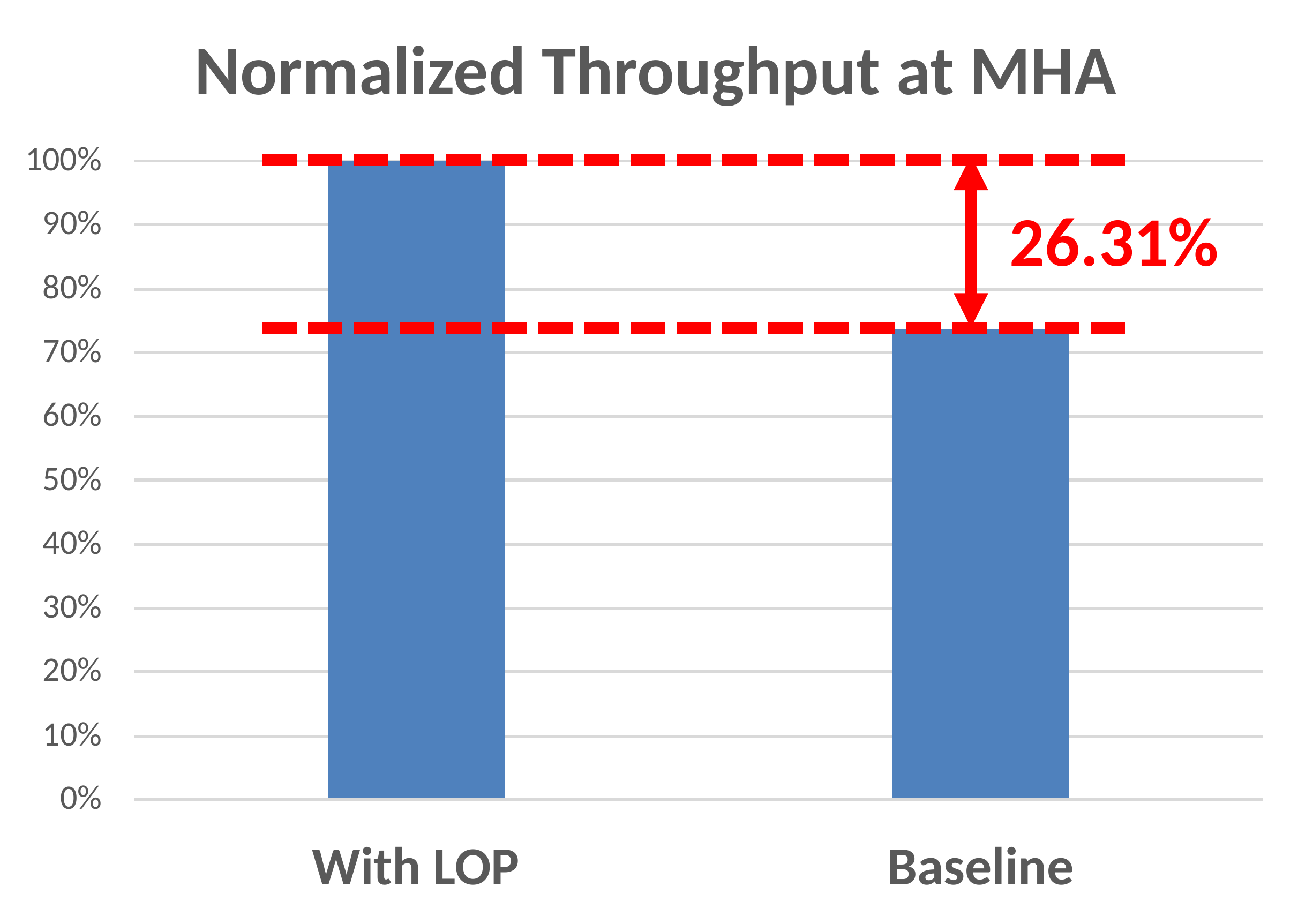}\hfill
  \includegraphics[width=0.46\linewidth]{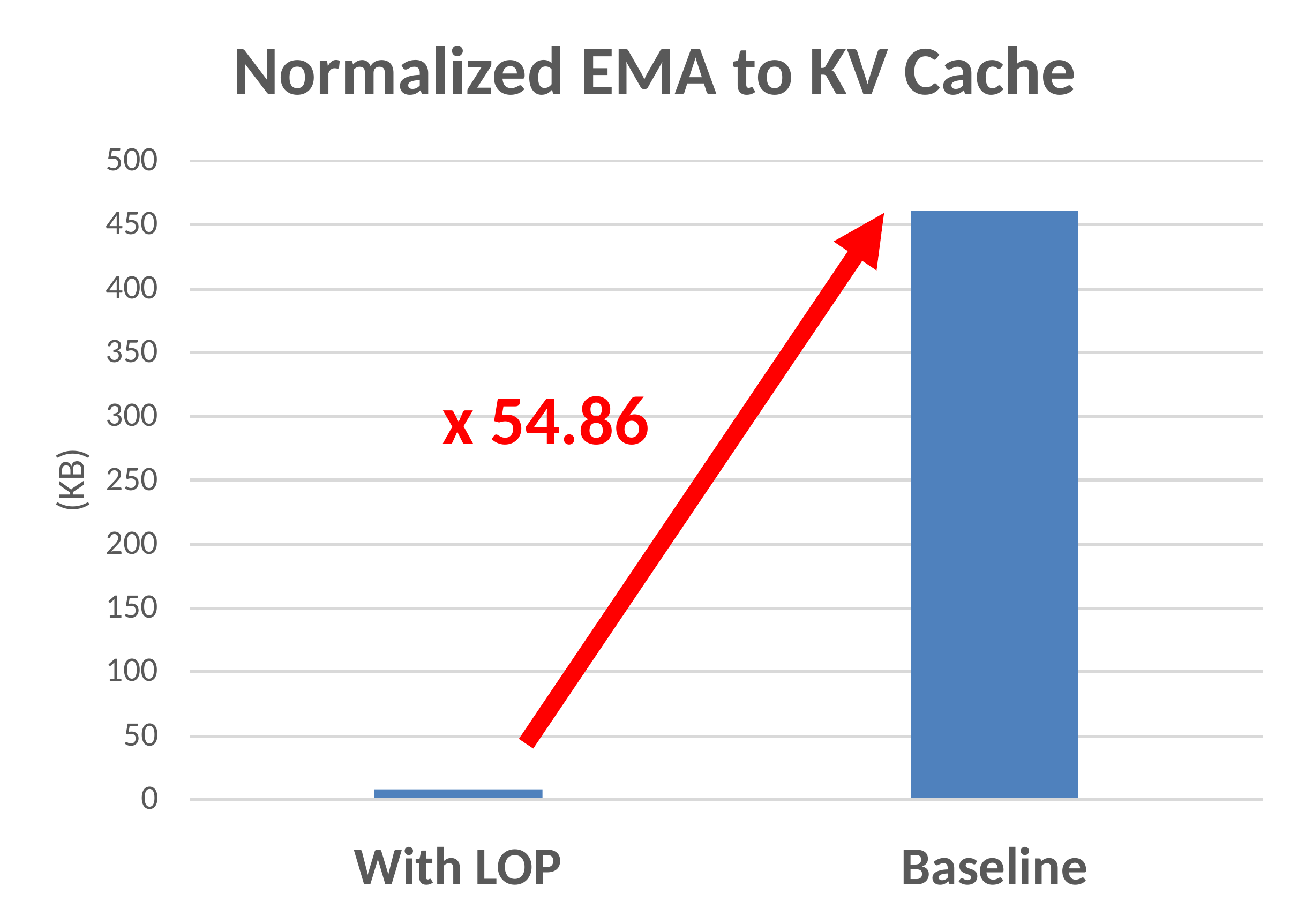}
  \vspace{-1mm}
  \caption{Effect of LOP: normalized MHA throughput (higher is better) and KV-cache EMA (lower is better).}
  \label{fig:abl_lop}
  \vspace{-1mm}
\end{figure}

\begin{figure}[htbp!]
  \centering
  \includegraphics[width=0.98\linewidth]{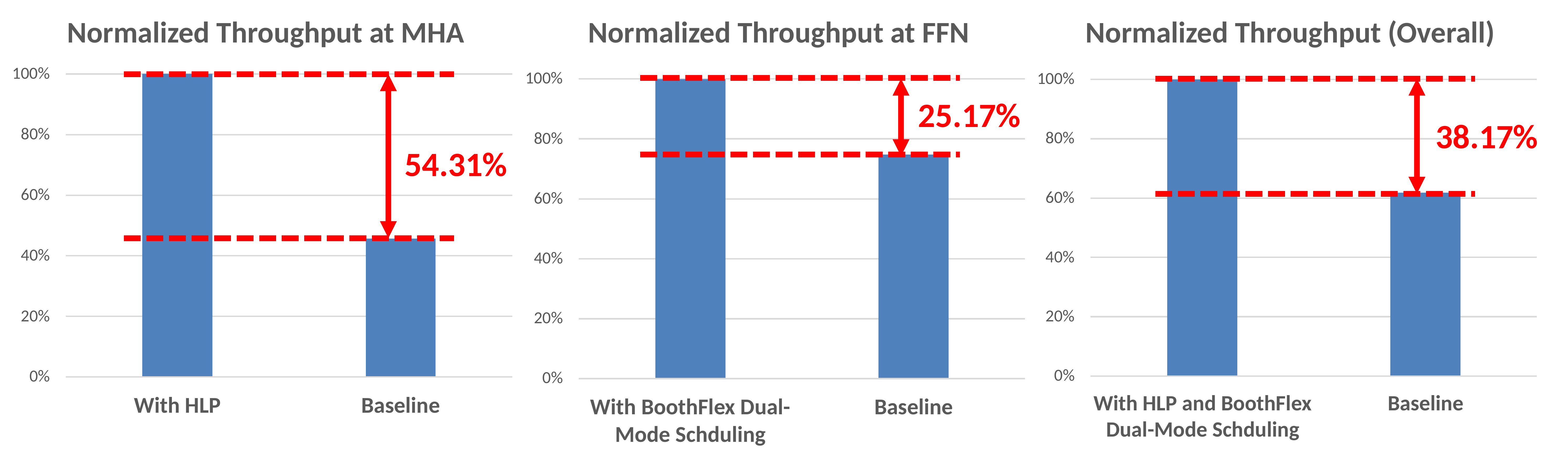}
  \vspace{-1mm}
  \caption{Throughput ablations: head-level pipelining (MHA), BoothFlex strategy (FFN), and overall with both enabled.}
  \label{fig:abl_hlp}
  \vspace{-1mm}
\end{figure}

\begin{itemize}
  \item \textbf{Predictive sparsity (LOP).} Relative to a baseline without LOP, MHA throughput rises by \textbf{26.31\%} and KV-cache traffic drops by \textbf{54.86$\times$} (Fig.~\ref{fig:abl_lop}), consistent with $(1{-}K/M)$—only $K$ candidates are fetched and computed.
  \item \textbf{Head-level pipelining (HLP).} Streaming at the head granularity overlaps \textbf{TINT} with \textbf{BoothFlex} for adjacent heads, yielding \textbf{+54.31\%} MHA throughput (Fig.~\ref{fig:abl_hlp}, left).
  \item \textbf{BoothFlex dual mode.} Enabling ternary–INT after MHA raises BoothFlex utilization from \textbf{0.51\%} to \textbf{69.20\%} (\textbf{$\times$135}) and improves FFN throughput by \textbf{25.17\%} (Fig.~\ref{fig:abl_hlp}, middle); combined with HLP, overall throughput increases by \textbf{38.17\%} (Fig.~\ref{fig:abl_hlp}, right).
\end{itemize}

%% file: chapters/5-Conclusion.tex
\section{Conclusion}
\label{sec:conclusion}

We presented \textbf{VitaLLM}, a mixed-precision edge accelerator that couples a multiplier-free \emph{TINT} core for ternary–INT projections with a \emph{BoothFlex} core that runs both INT8$\times$INT8 (attention) and ternary–INT (projections) on a shared Booth datapath, lifting utilization without a second array. A leading-one predictor with a comparison-free top-$K$ selector prunes KV fetches by roughly $1{-}K/M$ (for $M$ cached tokens) and restricts exact attention to only $K$ candidates, with no retraining. Per-head streaming consumes Q/K/V immediately, and an absmax-based barrier decouples nonlinear reductions (softmax/RMSNorm) from scaling, yielding a uniform \texttt{(integer vector, single scale)} interface across cores.

Our 16\,nm, 1\,GHz/0.8\,V prototype delivers \textbf{72.46\,tokens/s} decode within \textbf{0.214\,mm$^2$} at \textbf{59.12\,mW}. In ablations, LOP reduces KV traffic by \textbf{54.86$\times$} and speeds up MHA by \textbf{26.31\%}; head-level pipelining adds \textbf{54.31\%}; BoothFlex utilization rises to \textbf{69.20\%}, yielding an \textbf{overall +38.17\%} throughput increase when combined with HLP. These results indicate that VitaLLM enables practical 3B BitNet b1.58 inference on edge devices and offer a blueprint for efficient mixed-precision LLM accelerators.